\documentclass[10pt,letterpaper]{article}
\usepackage{hyperref}
\usepackage{graphicx}
\usepackage{float}
\usepackage[margin=1.4in]{geometry}
\usepackage[utf8]{inputenc}

\begin{document}

\title{A simple alternative to the Crystal Ball function}
\date{\today}
\author{Souvik Das\\ Department of Physics \\ University of Florida}
\maketitle

\begin{abstract}
We present a simple alternative to the Crystal Ball function that has an exponential tail stitched to a Gaussian core. It has one parameter less than the Crystal Ball function and, where appropriate, offers more stable fits to peaks that continue into exponential tails. The function may also be extended with two exponential tails on each side of the Gaussian, and this has two parameters less than the corresponding double-shouldered Crystal Ball function. This function has been used to model background and signal processes in a recent Higgs pair production search and may be of versatile use in experimental physics and other fields.
\end{abstract}

\bigskip

The Crystal Ball function, developed within the Crystal Ball Collaboration~\cite{Oreglia:1980cs, Skwarnicki:1986xj}, is a continuously differentiable ($C^1$) function that is often used as a fitting function in high energy physics. It is typically used to model lossy processes like the reconstructed invariant mass of a resonance from the energies and momenta of its decay products where some fraction of the energies and momenta are lost to detection. The Crystal Ball function consists of a power law tail stitched to a Gaussian core such that the function and its first derivative are continuous, as described in Eq.~\ref{eq:CrystalBall}. This results in 4 parameters: $\bar{x}$, $\sigma$, $\alpha$ and $n$. The power law parameter $n$ appears in the formula as $n^n$ and this sometimes makes the Crystal Ball an unstable fitting function. In this document, we report a simpler $C^1$ function that may be used to model similar peaks with long tails. It consists of an exponential tail stitched to the Gaussian core such that the function and its first derivative are continuous.

\begin{eqnarray}
\label{eq:CrystalBall}
f(x; \alpha, n, \bar{x}, \sigma) &=& e^{-\frac{1}{2}\left({\frac{x-\bar{x}}{\sigma}}\right)^2} \quad \textrm{for} \quad \frac{x-\bar{x}}{\sigma} > -\alpha \\
                                 &=& \left(\frac{n}{|\alpha|}\right)^n e^{-\frac{|\alpha|^2}{2}} \left( \frac{n}{|\alpha|} - |\alpha| - \frac{x-\bar{x}}{\sigma} \right)^{-n} \quad \textrm{for} \quad \frac{x-\bar{x}}{\sigma} \leq -\alpha \nonumber
\end{eqnarray} \\

The new function is described in Eq.~\ref{eq:GaussExp} with the exponential tail on the lower side of the Gaussian. $\bar{x}$ and $\sigma$ represent the mean and standard deviation of the Gaussian core, respectively. $k$ represents the decay constant of the exponential tail. Requiring continuity of the function and its first derivative implies that $k$ is also the number of standard deviations on the side of the tail where the Gaussian switches to the exponential. We call it the ``GaussExp" function.

\begin{eqnarray}
\label{eq:GaussExp}
f(x; \bar{x}, \sigma, k) &=& e^{-\frac{1}{2}\left(\frac{x-\bar{x}}{\sigma}\right)^2}, \quad \textrm{for} \quad \frac{x-\bar{x}}{\sigma} \geq -k  \\
                         &=& e^{\frac{k^2}{2}+k\left(\frac{x-\bar{x}}{\sigma}\right)}, \quad \textrm{for} \quad \frac{x-\bar{x}}{\sigma} < -k \nonumber
\end{eqnarray} \\

A similar function with exponential tails on both sides of the Gaussian core is expressed in Eq.~\ref{eq:ExpGaussExp}. $k_L$ and $k_H$ are the decay constants of the exponentials on the low and high side tails. We call it the double-shouldered GaussExp, or the ``ExpGaussExp" function.

\begin{eqnarray}
\label{eq:ExpGaussExp}
f(x; \bar{x}, \sigma, k_L, k_H) &=& e^{\frac{k_L^2}{2}+k_L\left(\frac{x-\bar{x}}{\sigma}\right)}, \quad \textrm{for} \quad \frac{x-\bar{x}}{\sigma} \leq -k_L \\
                         &=& e^{-\frac{1}{2}\left(\frac{x-\bar{x}}{\sigma}\right)^2}, \quad \textrm{for} \quad -k_L < \frac{x-\bar{x}}{\sigma} \leq k_H  \nonumber \\
                         &=& e^{\frac{k_H^2}{2}-k_H\left(\frac{x-\bar{x}}{\sigma}\right)}, \quad \textrm{for} \quad k_H < \frac{x-\bar{x}}{\sigma} \nonumber
\end{eqnarray} \\

A recent CMS paper on the search for resonant pair production of Higgs bosons decaying to two bottom quark-antiquark pairs~\cite{Khachatryan2015560} has used these functions to model the invariant masses of background and signal processes. The GaussExp function with a high-side tail was used to model the invariant mass of the background which consists of four candidate jets presumably originating from bottom quarks and passes the event selection criteria of the analysis. The invariant mass of the signal was modeled with the ExpGaussExp function.

\section{Example Fits}

This section contains an example of an exponentially falling distribution fit with the Crystal Ball function and compared to a fit with the GaussExp function. It also contains an example of a simulated $H\to\gamma\gamma$ peak fit with the ExpGaussExp function and compared to a fit with the double-shouldered Crystal Ball function. In both cases, the new functions perform at least as well as the Crystal Ball function with less fit parameters and without instabilities related to the Crystal Ball exponent $n$. These functions may therefore find more versatile use in experimental physics and other fields.

\subsection{An exponentially falling distribution with a turn-on region}

\begin{figure}[H]
\centering
\includegraphics[width=0.49\textwidth]{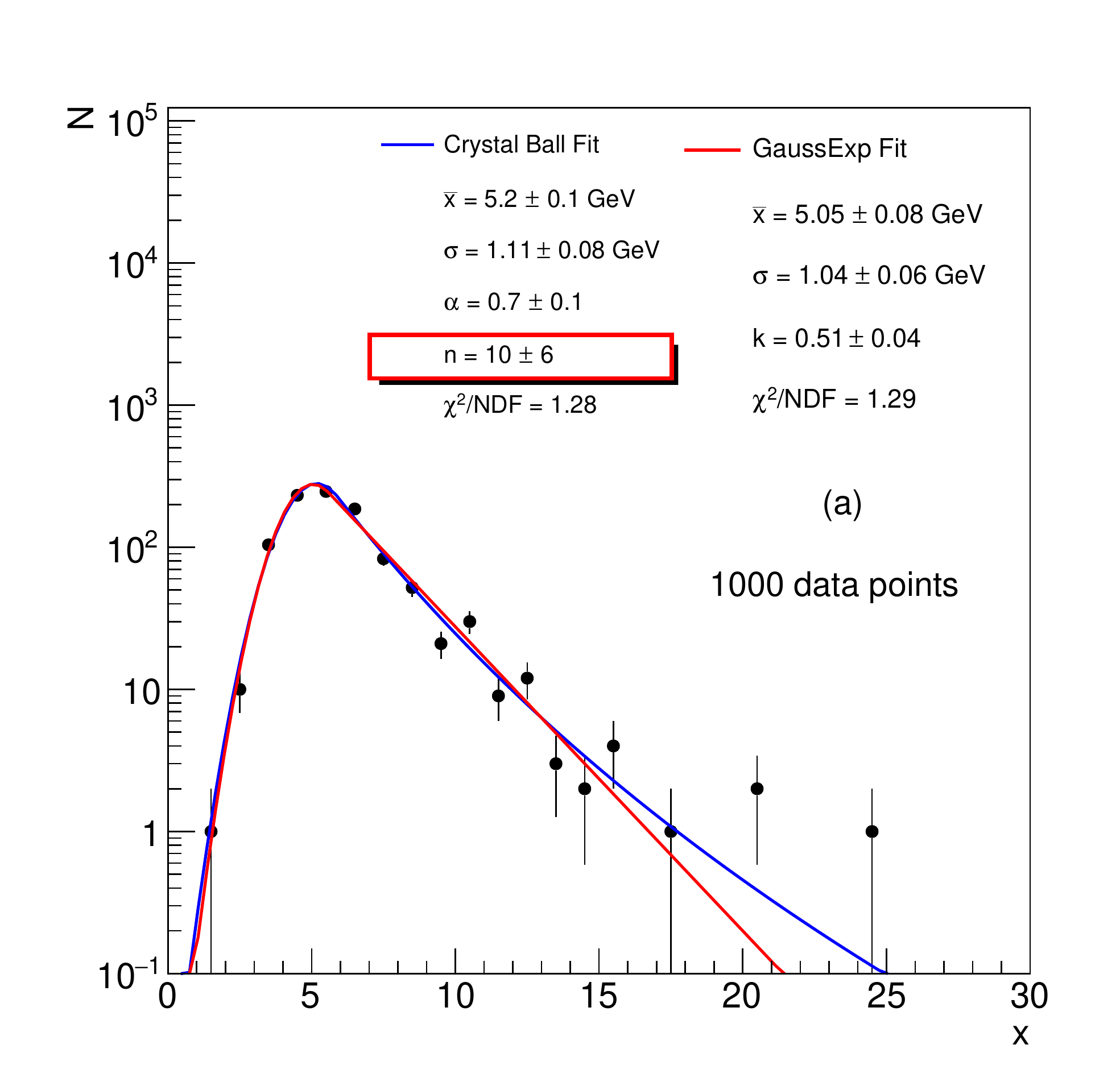}
\includegraphics[width=0.49\textwidth]{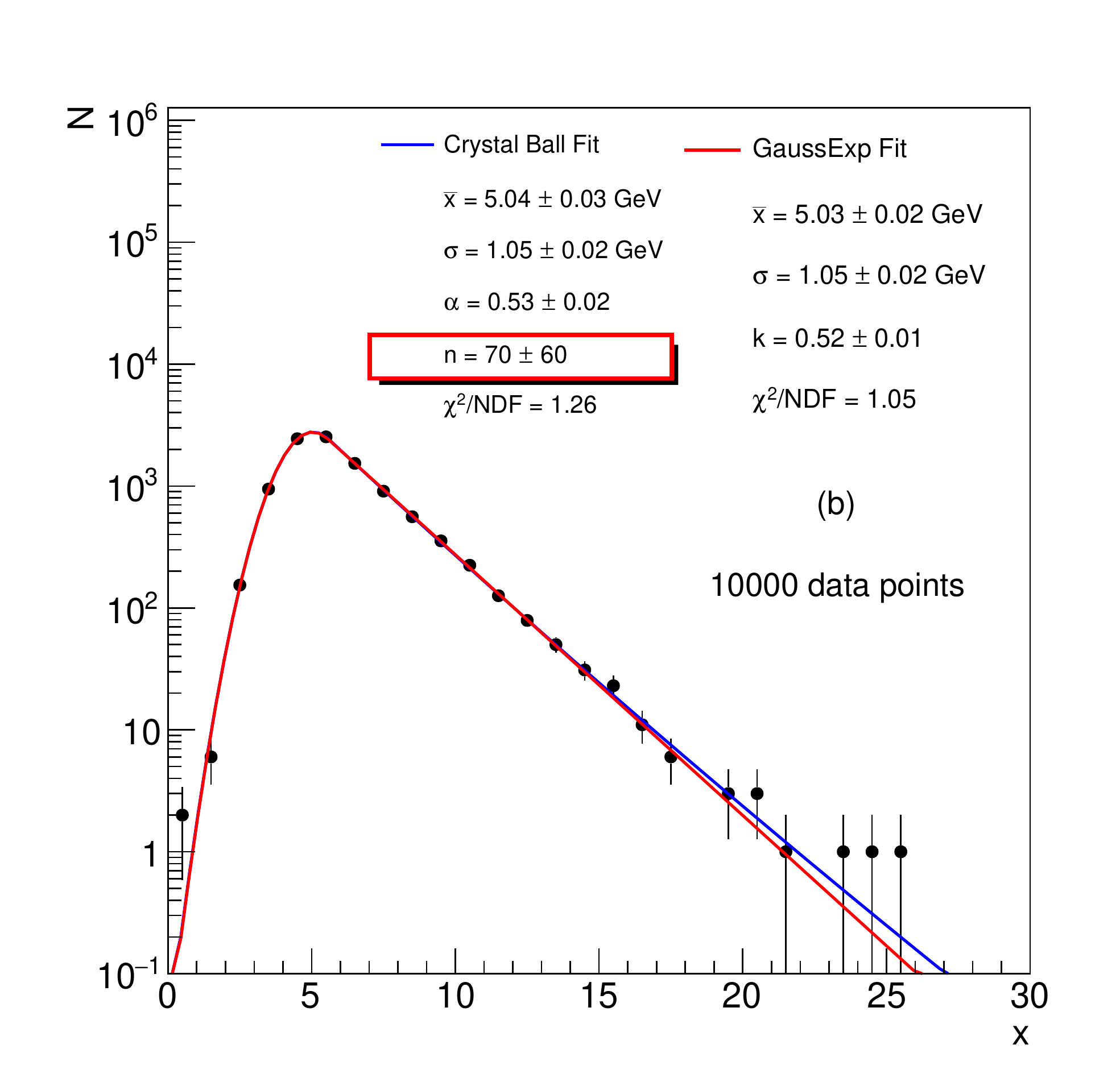}
\includegraphics[width=0.49\textwidth]{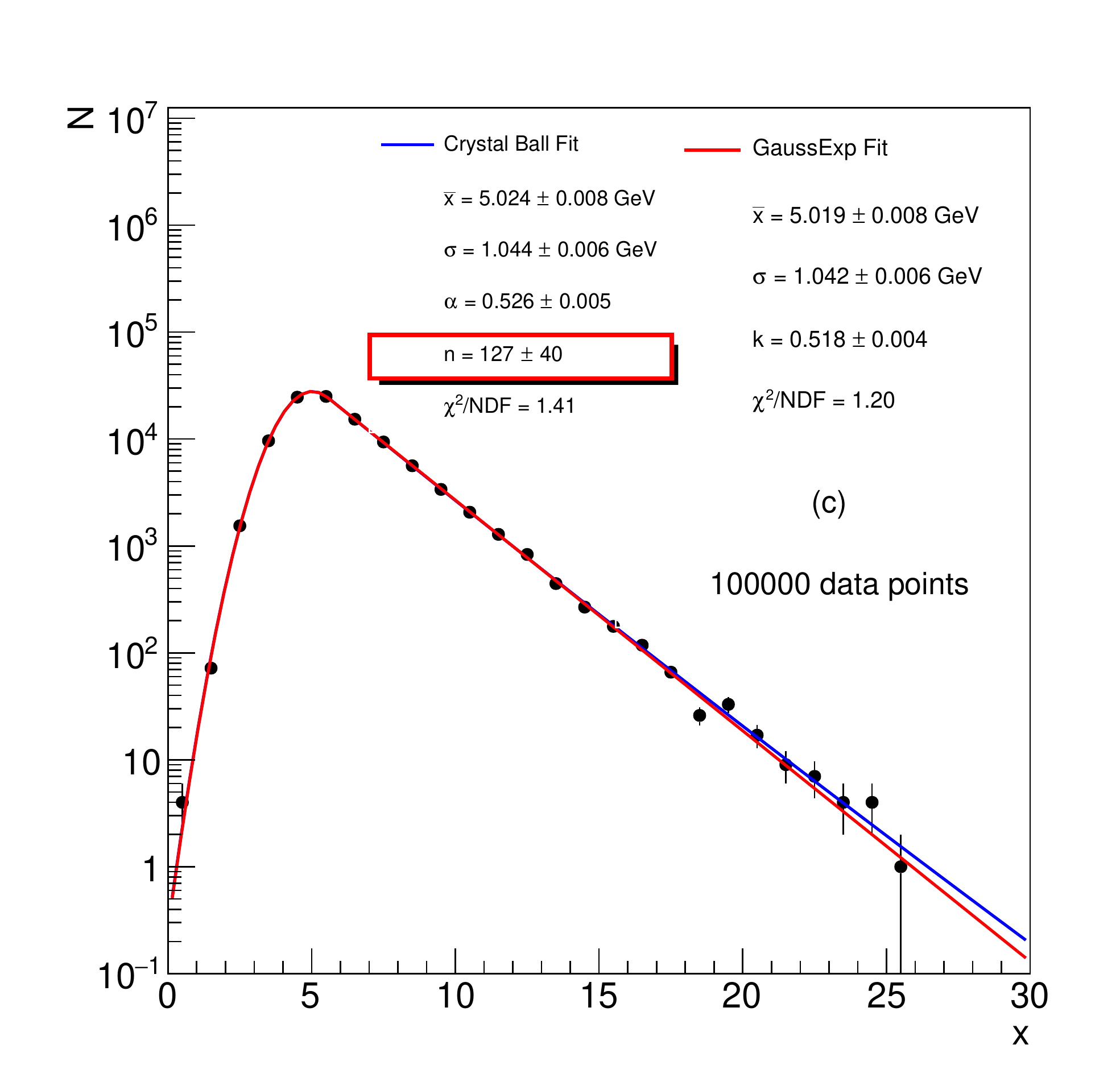}
\includegraphics[width=0.49\textwidth]{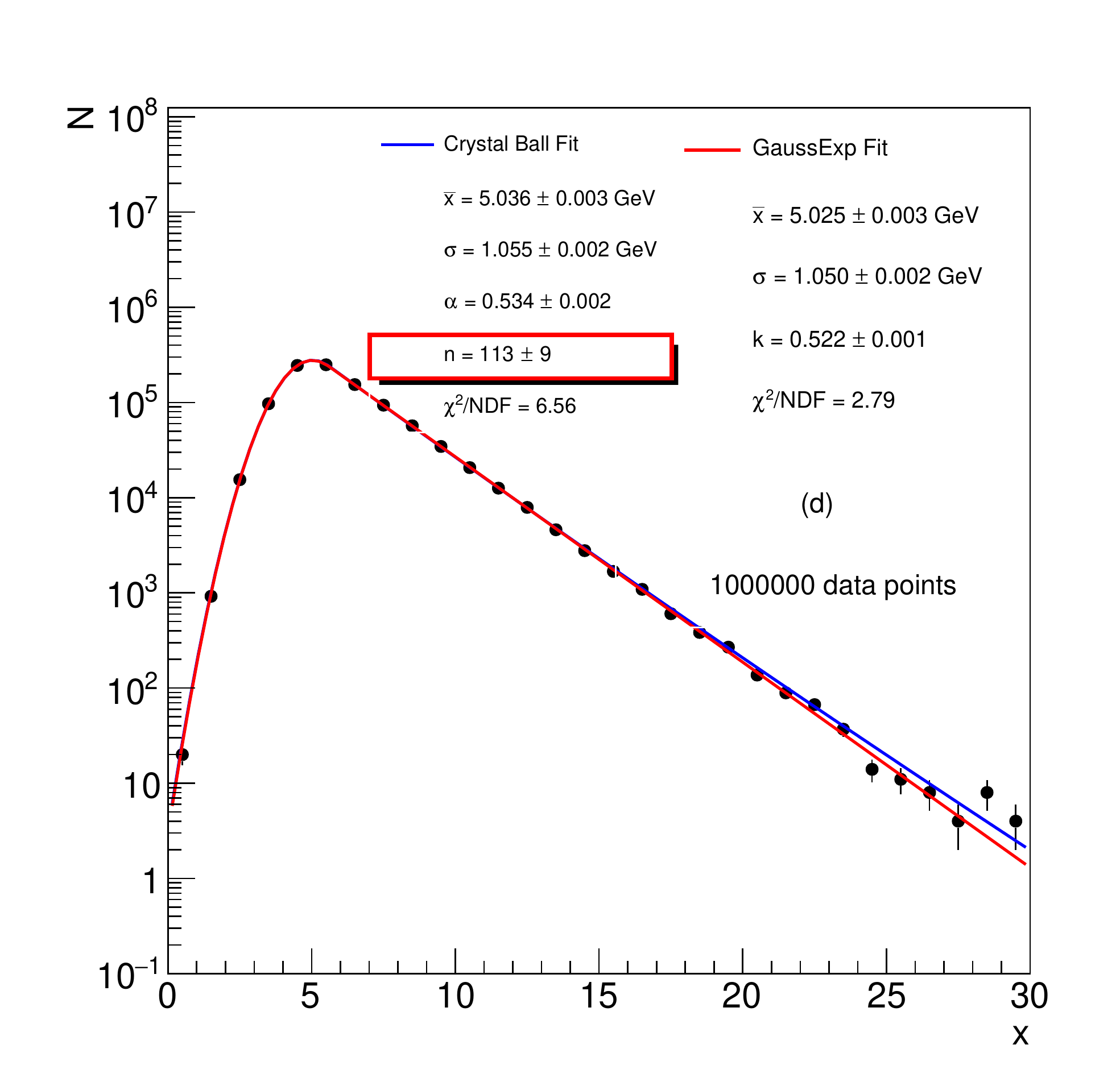}
\caption{(a) 1,000, (b) 10,000, (c) 100,000 and (d) 1,000,000 random data points thrown from the GaussExp distribution and fitted to both Crystal Ball (blue line) and GaussExp (red line) functions with maximum likelihood. The Crystal Ball fits converge in (a) and (b) but with large uncertainties in $n$, and fail to converge in (c) and (d).}
\label{fig:ThrowCatch}
\end{figure}

There may occur natural processes that are characterized in some variable with a turn-on region that merges into an exponentially falling tail. If the turn-on region is sufficiently well modeled with a Gaussian, modeling the tail with a power-law would not be appropriate and would result in unstable fits and uncertainties on the parameters of the fit. We demonstrate this here with 1,000, 10,000, 100,000 and 1,000,000 random data points thrown within 0 and 30 from the GaussExp function with $\bar{x} =$ 5, $\sigma =$ 1 and $k =$ 0.5. Each of these distributions shown in Fig.~\ref{fig:ThrowCatch} are fitted by maximum likelihood to both the Crystal Ball function and the GaussExp function using the MINUIT fitting program~\cite{JAMES1975343} within the ROOT framework~\cite{Antcheva20092499}.

With 1,000 throws shown in Fig.~\ref{fig:ThrowCatch}a, MINUIT converges with both the Crystal Ball and GaussExp fits and reports a similar $\chi^2/NDF$. The Crystal Ball exponent $n$, however, is returned with a 60\% uncertainty. With 10,000 throws shown in Fig.~\ref{fig:ThrowCatch}b, the fitter still converges with the Crystal Ball function but reports a worse $\chi^2/NDF$ than the fit with the GaussExp function. $n$ is returned with an 86\% uncertainty and this indicates that the Crystal Ball function is poorly constrained. With 100,000 throws shown in Fig.~\ref{fig:ThrowCatch}c, the fits of the Crystal Ball and GaussExp functions look similar by eye but the fitter reports a failure with the Crystal Ball function. The Crystal Ball $n$ is returned as 127 with an uncertainty of 40. Expressions like $n^n$ in Eq.~\ref{eq:CrystalBall} for the Crystal Ball function make for an irregular likelihood function for high values of $n$ and this is the reason the fitter fails to find a stable maximum. With 1,000,000 throws shown in Fig.~\ref{fig:ThrowCatch}d, the fitter fails with the Crystal Ball function. The $n$ is returned as 113 $\pm$ 9. A fit parameter that depends so sensitively on the number of events in the distribution cannot be said to characterize the distribution.

The GaussExp fit, on the other hand, converges every time and returns fit parameters with small uncertainties that are consistent with the parameters of the parent distribution in these examples.

\subsection{A signal peak}

\begin{figure}[H]
\centering
\includegraphics[width=0.49\textwidth]{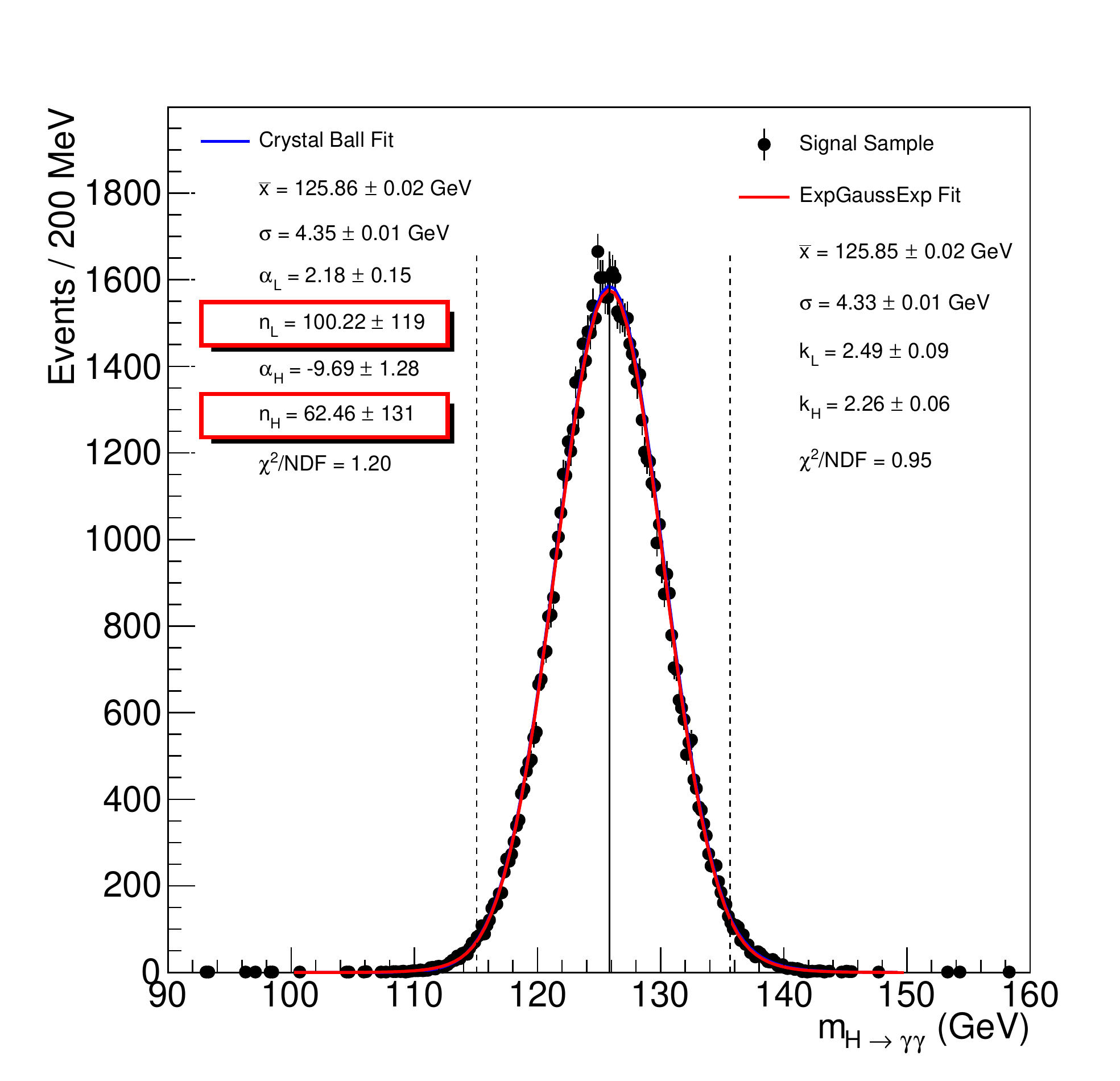}
\includegraphics[width=0.49\textwidth]{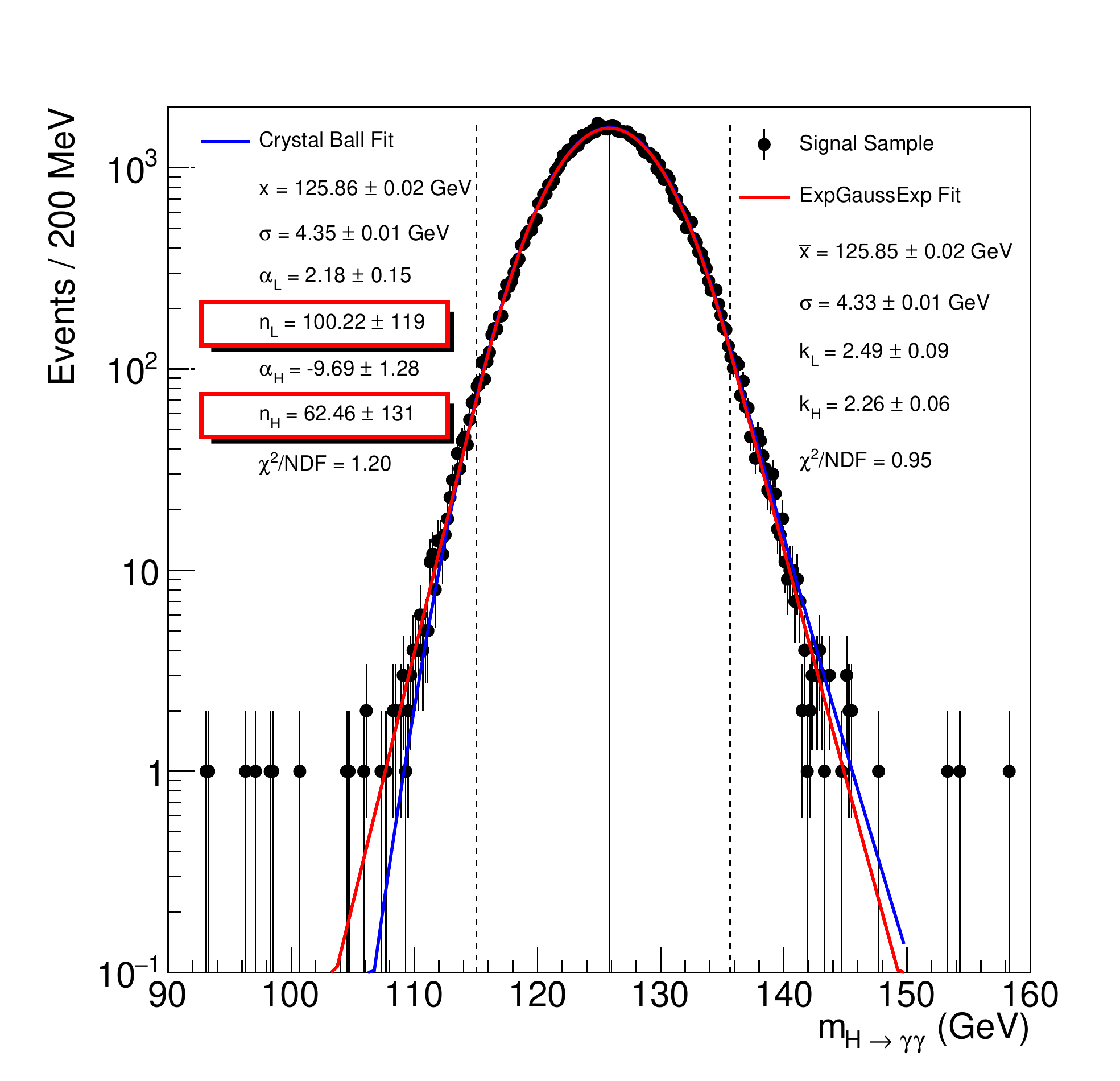}
\caption{Invariant mass of the Higgs boson in simulated $H\to\gamma\gamma$ processes fit to the ExpGaussExp function (red line) with maximum likelihood. The plot on the left shows the fit on a linear scale. The plot on the right shows the fit on a logarithmic scale. The solid vertical line indicates the mean of the Gaussian core. The dashed vertical lines indicate the switch to exponential decays. A corresponding Crystal Ball fit is also shown (blue line).}
\label{fig:HGGFit}
\end{figure}

We simulate the $H\to\gamma\gamma$ process in 13 TeV proton-proton collisions in Madgraph~\cite{Alwall:2011uj}, shower and hadronize events in Pythia~\cite{PYTHIA}, and reconstruct events from a parametrized simulation of the CMS detector response in Delphes~\cite{deFavereau:2013fsa} to obtain a sample for fitting here. The two highest energy photons are combined in these simulated events to reconstruct the Higgs invariant mass peak as shown in Fig.~\ref{fig:HGGFit}. The detector response produces tails that cannot be fit with a simple Gaussian.

We fit this distribution with maximum likelihood to the double-shouldered Crystal Ball and the ExpGaussExp functions. These fits are performed in the 100 to 150 GeV range. MINUIT reports not just a higher $\chi^2/NDF$ with the Crystal Ball function but also a failure to fit with it. This can be traced back to the exponents $n_L$ and $n_H$ which are returned as 100 $\pm$ 120 and 63 $\pm$ 131, respectively. Large values of these exponents imply large derivatives in the likelihood function and cause the fitter to fail. On the other hand, the ExpGaussExp fit is successful, reports a $\chi^2/NDF$ close to 1 and very small uncertainties in all its parameters.

\section*{Acknowledgments}

We thank Dr. Saptaparna Bhattacharya for generating the $H\to\gamma\gamma$ signal sample using Madgraph, Pythia and Delphes, and for helpful discussions on the use of the GaussExp and ExpGaussExp functions. We also thank Dr. Jacobo Konigsberg for further helpful discussions in making the case for these new functions.

\bibliographystyle{plain}
\bibliography{GaussExp}

\end{document}